\documentclass[intlimits,twoside,a4paper]{article}

\usepackage[eqsecnum]{cmpj3}

\usepackage{bm}

\issue{2024}{27}{1}{13802}
\doinumber{10.5488/CMP.27.13802}
\title[Multilayered  bacterial colonies]%
{Toward a realistic model of multilayered bacterial colonies%
}  
\author[M. T. Khan, J. Cammann, A. Sengupta, E. Renzi, M. G. Mazza]{M. T. Khan\orcid{0009-0004-8831-7359}\refaddr{label1}\thanks{Corresponding author: \email{m.g.mazza@lboro.ac.uk}.},
	J. Cammann\orcid{0000-0003-3245-8078}\refaddr{label1},
	A. Sengupta\orcid{0000-0001-5592-7864}\refaddr{label2,label3},
	E. Renzi\orcid{0000-0002-1459-5565}\refaddr{label4},
	M. G. Mazza\orcid{0000-0002-5625-9121}\refaddr{label1}
}
\addresses{
	\addr{label1} Interdisciplinary Centre for Mathematical Modelling and Department of Mathematical Sciences, Loughborough University, Loughborough, Leicestershire LE11 3TU, United Kingdom
	\addr{label2} Physics of Living Matter Group, Department of Physics and Materials Science, University of Luxembourg, 162 A, Avenue de la Fa\"iencerie, L-1511 Luxembourg City, Luxembourg
	\addr{label3} Institute for Advanced Studies, University of Luxembourg, 2 Avenue de l'Universit\'e, L-4365 Esch-sur-Alzette, Luxembourg
	\addr{label4} Mathematics of Complex and Nonlinear Phenomena (MCNP), Department of Mathematics, Physics and Electrical Engineering, Northumbria University, Newcastle upon Tyne, NE1 8ST, United Kingdom
}


\date{Received January 12, 2024, in final form February 16, 2024}

\begin{document}
	
	\maketitle
	
	\begin{abstract}
		Bacteria are prolific at colonizing diverse surfaces under a widerange of environmental conditions, and exhibit fascinating examples of self-organization across scales. Though it has recently attracted considerable
		interest, the role of mechanical forces in the collective behavior of bacterial colonies is not yet fully understood. Here, we construct a model of growing rod-like bacteria, such as \textit{Escherichia coli} based purely on mechanical forces. We perform overdamped molecular dynamics simulations of the colony starting from a few cells in contact with a surface. As the colony grows, microdomains of strongly aligned cells grow and proliferate. Our model captures both the initial growth of a bacterial colony and also shows characteristic signs of capturing the experimentally observed transition to multilayered colonies over longer timescales.  We compare our results with experiments on \textit{E.~coli} cells and analyze the statistics of microdomains. 
		\printkeywords
	\end{abstract}

	\section{Introduction}

	Microbial life in general exists not as cells freely floating in an aqueous environment (planktonic state), but rather as sessile, surface associated colonies \cite{costerton1995microbial}. These sessile, self-organized communities often develop into biofilms, defined as matrix-enclosed colonies adhering to each other and/or to surfaces~\cite{costerton1995microbial,hartmann2019emergence}. This matrix, called extracellular polymeric substances (EPS), is composed of biopolymers, nucleic acids, and lipids which produce cohesion of the biofilm, adhesion to the substrate, and a number of protective functions \cite{flemming2010biofilm}.
	Starting from the initial attachment, followed by the growth, proliferation, and ultimately colonization, biofilms represent self-regulated active microbial communities where an interplay of geometry, order and growth-mediated mechanics determine their emerging structure and topology \cite{araujo2023steer, nijjer2021mechanical,sengupta2020topology,allen2018bacterial,drescher2016architectural,karimi2015interplay,farrell2013mechanically}. From dental surfaces and open wounds,  cancer and gut environments to the biological growth observed on maritime structures (biofouling), biofilms can grow and thrive under diverse settings, thereby regulating infectious diseases and human health, as well as determining the durability of maritime vessels \cite{tiron2015biofouling} and medical devices such as  intravenous catheters, prosthetic heart valves, joint prostheses, peritoneal dialysis catheters, and cardiac pacemakers \cite{hall2004bacterial}. Importantly, biofilms can also impact the respiratory mucous and the airways of patients with cystic fibrosis, thus regulating patient health and response to medical treatment \cite{mazza2016physics}.
	Because of the specificity of the physical and biological interactions among its constituent cells, biofilms are often considered an emergent form of microbial life, as the properties and function of biofilms may outperform those of the constituent cells \cite{hall2004bacterial}.

	The first step in the development of a biofilm is the growth of a colony of cells multiplying on a substrate, initially at an exponential rate, until nutrient limitations cause a plateauing of the colony size. In this initial phase of rapid expansion, biofilms self-organize into monolayers wherein contact forces between neighboring cells and with the substrate play a prominent role \cite{you2018geometry,you2019mono}.
	These contact forces lead to an orientational transition reminiscent of liquid-crystalline phase transitions. As the degree of confinement decreases, rod-like cells transition from a long-ranged nematic state to a globally disordered state characterized by microdomains with short-range nematic order \cite{dell2018growing,boyer2011buckling,sheats2017role,you2018geometry}.

	Considerable work has focused on the transition from a two-dimensional monolayer of cells to a three-dimensional (3D) biofilm. A crucial step in the development of a 3D structure is the verticalization of cells \cite{beroz2018verticalization,you2019mono}, where the growth of mechanical stresses among the growing cells leads to a buckling instability that turns some cells away from a planar configuration parallel to the substrate \cite{you2019mono}.
	
	Recent work has shown that, following verticalization, the second layer spreads over the first layer, and thereafter the third layer starts to develop on the second layer and so on \cite{dhar2022self}. Over time, this \textit{monolayer stacking} process leads to the 3D morphology of the bacterial colony emerging from multiple flat layers of dividing cells parallel to the underlying flat substrate.
	Developing a model for this monolayer
	stacking transition is the main goal of this work.

	\section{Bacterial model}
	
	\begin{figure}[!t]
		\centering
		\includegraphics[width=0.30\columnwidth]{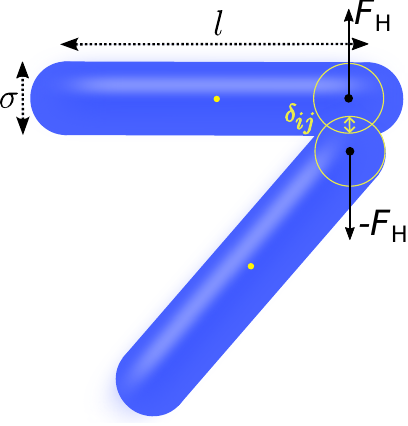}
		\caption{(Colour online) Schematic of two spherocylinders representing rod-like bacteria (e.g.,~\textit{E. coli} cells) of  length $l$ (excluding the hemispherical caps) and diameter $\sigma$. Thin yellow circles represent the virtual spheres used to compute the Hertzian repulsion $\bm{F}_\text{H}$ based on the overlap of the spheres $\delta_{ij}$.}
		\label{fig:Interaction between rods}
	\end{figure}

	We model each bacterial cell as a spherocylinder \cite{earl2001computer} of length $l(t)$, depending on time $t$, and diameter $\sigma$ capped at both ends by hemispheres of the same diameter $\sigma$. Each cell is described by the vector to its center-of-mass position  $\bm{r}$ and a unit vector for its orientation $\bm{\hat{e}}$.
	In the overdamped limit, the equations of motion read
	\begin{align}
		\frac{\rd\bm{r}}{\rd t} = & \frac{1}{\Gamma} \left(\bm{F}_\text{H}+\bm{F}_\text{Hs}-\frac{\partial U_\text{cc}}{\partial\bm{r}} - \frac{\partial U_\text{cs}} {\partial\bm{r}} + \bm{g}\right)  , 
		\label{eqn: Equation of motion in space} \\
		\frac{\rd\bm{\hat{e}}}{\rd t} = & \frac{1}{\Omega}({I - \bm{\hat{e}} \otimes\bm{\hat{e}}}) \left( \bm{\tau}_\text{H}-\frac{\partial U_\text{cc}}{\partial\bm{\hat{e}}} - \frac{\partial U_\text{cs}}{\partial\bm{\hat{e}}} + \bm{\tau} \right),
		\label{eqn: Equation of motion for orientation}
	\end{align}
	where $\Gamma$ and $\Omega$ are characteristic translational and rotational drag coefficient, respectively,  $I$ is the identity matrix \cite{pearce2019flow}. In equations~\eqref{eqn: Equation of motion in space}--\eqref{eqn: Equation of motion for orientation} $\bm{g}=-0.5 \bm{\hat{z}}$ represents a constant adhesion force applied to center of each rod and $\bm{\tau}=-(l+\sigma)(\bm{\hat{e}}\cdot \bm{\hat{z}}) \bm{\hat{z}}$ is a restoring torque that returns our rods to planar alignment, respectively, and $\bm{\hat{z}}$ is the  unit vector normal to the substrate. 
	Both  $\bm{g}$ and $\bm{\tau}$ physically stem from the adhesive forces within the EPS matrix produced by and surrounding the cells. Additionally, to model thermal fluctuations, we include a small fluctuating torque with amplitude $\propto 10^{-7}$.

	In equation~\eqref{eqn: Equation of motion in space}, $\bm{F}_\text{H}=E_0\sigma^{1/2}\delta^\frac{3}{2}_{ij} \bm{n}_{ij}$, where $\bm{n}_{ij} \equiv (\bm{r}_i-\bm{r}_j)/r_{ij}$, and $r_{ij}=|\bm{r}_i-\bm{r}_j|$ is the distance between the center of the cells; $\bm{F}_\text{H}$ represents a Hertzian repulsion force 
	between overlapping spherocylinders with cell stiffness  $E_0 = 100$ (dependent on Young's modulus) \cite{volfson2008biomechanical,beroz2018verticalization}; this repulsion is calculated by means of virtual spheres of diameter $\sigma$ centered at the closest points between the axes of spherocylinders so that the cylinders are in contact whenever the distance between the centers of the virtual spheres $\bar{r}_{ij}$ is such that $\bar{r}_{ij}-\sigma \equiv\delta_{ij}<0$.
	The smallest distance between two spherocylinders is computed using the algorithm by Vega and Lago \cite{vega1994fast}. The substrate also exerts a Hertzian repulsion $\bm{F}_\text{Hs}=E_0\sigma^{1/2}\delta^\frac{3}{2}_{i\text{s}} \bm{\hat{z}}$, where $\delta_{i\text{s}}$ is the overlap of the spherocylinder with the solid substrate. In equation~\eqref{eqn: Equation of motion for orientation}, $\bm{\tau}_\text{H}=\sum (\bm{r}_\text{c}-\bm{r}) \times \bm{F}_\text{H}$, where $\bm{r}_\text{c}$ is the vector to the contact point. 
	
	We use the following potentials to model the cell-cell interaction 
	$U_\text{cc}$ acting between each bacterium and their neighboring cells and the cell-substrate potential $U_\text{cs}$ with the confining surface \cite{pearce2019flow}
	\begin{equation}
		U_\text{cc} =  \varepsilon_\text{cc} \frac{ [1+(\bm{\hat{e}}_i\cdot \bm{\hat{e}}_j)^2] }{1 + \exp{\frac{\rho-r_{ij}}{\lambda}}}\,, \quad  U_\text{cs} = \varepsilon_\text{cs}  \frac{\exp(\gamma (h-h_0))}{h-h_0}\,.
		\label{eqn:cell-cell_potential}
	\end{equation}

	%
	%
	In equation~\eqref{eqn:cell-cell_potential}, $U_\text{cc}$ represents a short-range cell-cell attraction with relative strength $\varepsilon_\text{cc}  = -10^{-3}$,  where $\rho = 2.83\sigma$ is the attraction position, $\lambda = 0.16 \sigma$ the width of the potential  \cite{pearce2019flow},
	%
	whereas $U_\text{cs}$ represents the attraction between cell and substrate modelled through a Yukawa-like potential with $\varepsilon_\text{cs}  = -2.5$, where $\gamma = 1$ is the depth of Yukawa potential,  $h$ is the relative distance between cell and substrate, and is $h_0$  the position of the substrate in our simulations.

	\begin{figure}[!t]
		\centering
		\includegraphics[width=\linewidth]{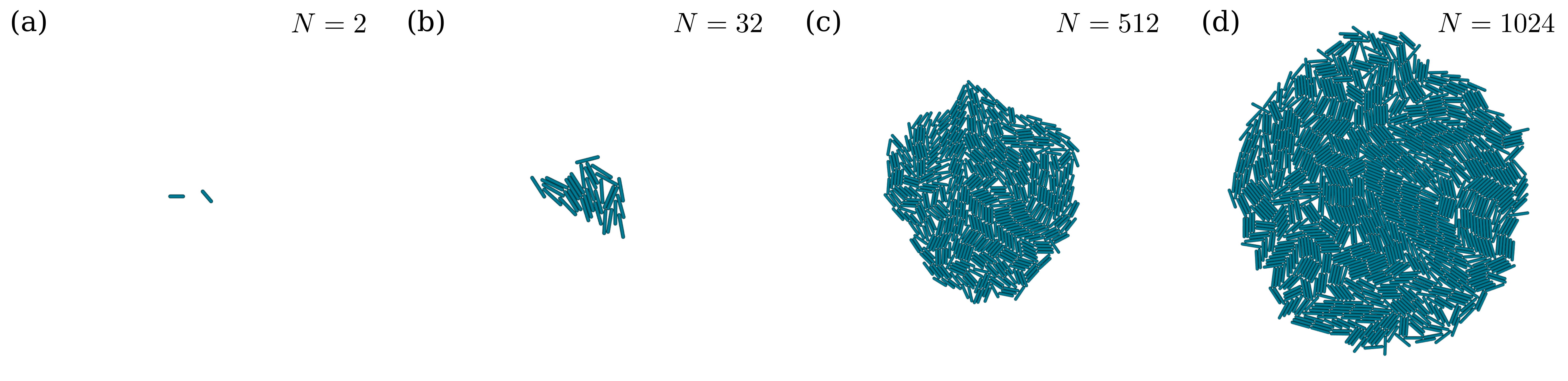}
		\caption{(Colour online) Different stages of the colony's life. Starting from an initial condition with two cells randomly oriented (a), the cells undergo different cycles of growth and division (b) and then (c). Eventually, the colony assumes a circular shape with most cells at its boundary aligned tangentially to the colony (d). Each snapshot indicates the number of cells $N$ present. The underlying substrate is not shown.}
		\label{fig:4-pane_growth}
	\end{figure}

	Bacterial cell proliferation is the driving force underpinning  mechanical interactions in a colony~\cite{wittmann2023collective,willis2017sizing,jun2018fundamental}. 
	Away from division events, the cell length growth obeys a linear law (with coefficient $G_R$) modified to include a mechanoresponse $\phi(\delta)$ of the cell to the growing pressure exerted by other cells due to proliferation stresses 
	\begin{equation}
		\frac{\rd l}{\rd t} = G_R \,\phi(\delta) \,, \quad \phi(\delta)=\left[ 1- \frac{ \alpha  (\delta - \delta_0)}{1 + \alpha |\delta - \delta_0| } \right]\,. 
		\label{eqn: growth rate}
	\end{equation}
	A cell divides when it reaches a split length $l_s=6\sigma$, a small random offset in the daughter
	cells' length is also used. 
	In equation~\eqref{eqn: growth rate}, $G_R = 10^{-7}$ is a growth rate constant.  It has been observed that when colonies grow, the increasing pressure exerted by each cell on its neighbors leads to a slowdown in the cell length growth, akin to a contact inhibition \cite{volfson2008biomechanical,wittmann2023collective}; the term in square brackets is a sigmoid function that models such mechanoresponse to the growing pressure, where $\alpha=60$ represents a scale parameter regulating the steepness of this mechanoresponse, $\delta$ is the total overlap between a cell and its neighbors, and $\delta_0=0.12 \sigma$ is the inflection point of this dependence.
	To avoid artificial synchronization effects in cell divisions across the colony, when a cell divides, the two daughter cells differ in length by a random amount with $5\%$ maximum amplitude. 
	
	Figure \ref{fig:Interaction between rods} shows a schematic illustration of two spherocylinders; the thin yellow circles represent the virtual spheres used to compute the overlap and the Hertzian repulsion $\bm{F}_\text{H}$  between the bacterial cells.
	We perform molecular dynamics simulations using an Euler integration scheme with a time step $\Delta t=10^{-5}$.

	\section{Results and discussion}

	\begin{figure}[!t]
		\centering
		\includegraphics[width=0.6\linewidth]{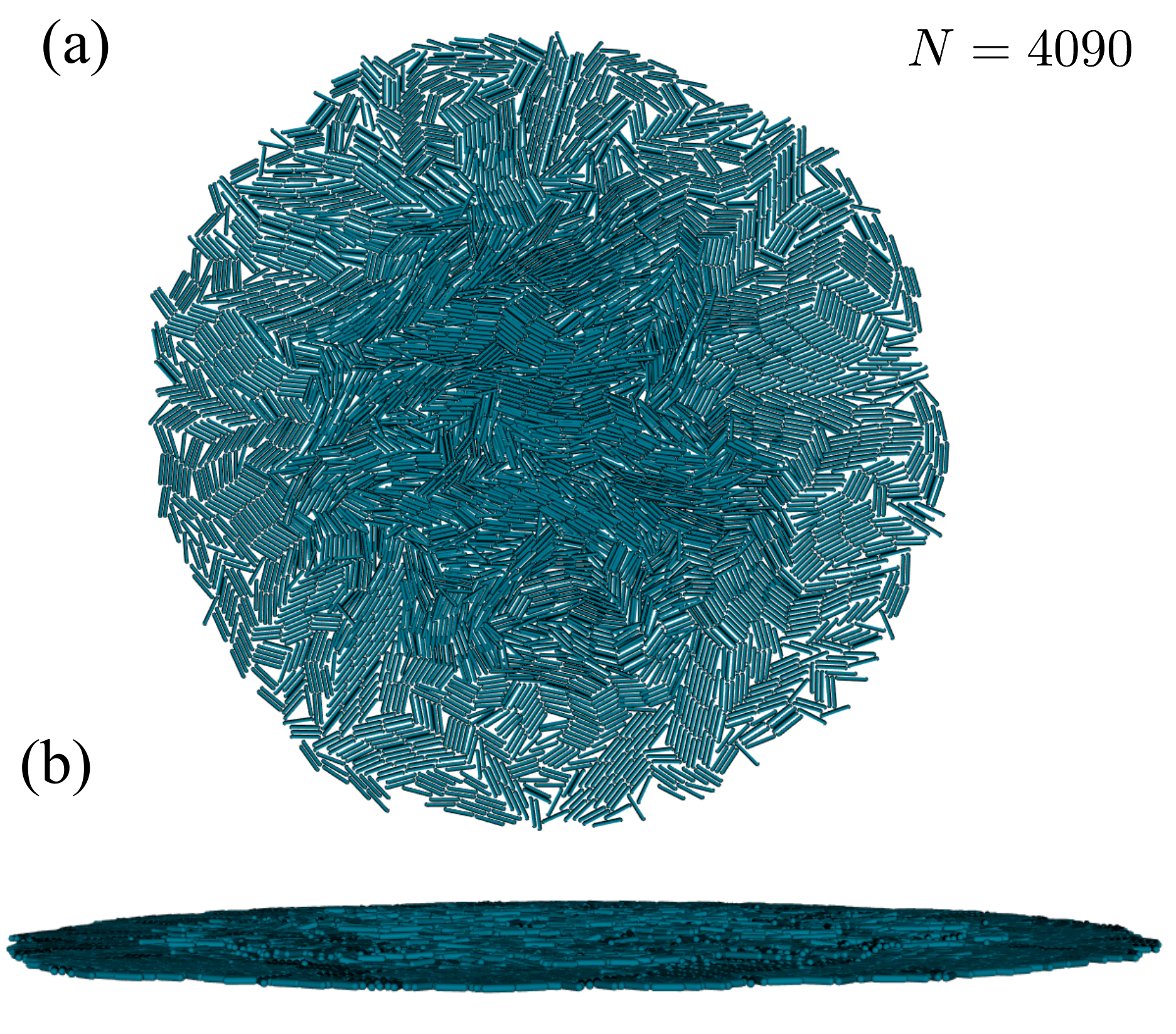}
		\caption{(Colour online) (a) Top view of a large bacterial colony which underwent mono to multiplayer transition. The second layer of cells is visible as the darker region in the center. (b) Edge view of the same colony showing the presence of extruded cells which have divided to form a flat second layer. The number of cells $N$ is indicated in the snapshot.}
		\label{fig:two_layer_view}
	\end{figure}

	Figure \ref{fig:4-pane_growth} shows a sequence of snapshots from a simulated
	colony growing from a starting configuration of two randomly oriented cells that, after a number of cycles of growth and division reaches a maturation stage comprising multiple layers \cite{you2019mono}. In agreement with experimental evidence \cite{su2012bacterial,grant2014role}, the colony initially grows as a flat, circular monolayer. We note that our simulations reproduce the tangential alignment of cells at the outer edge of the colony. This is due to the fact that a tangentially-oriented rod is mechanically stable to the forces of neighboring cells, whose torques cancel out statistically \cite{su2012bacterial}.
	
	\begin{figure}[!t]
		\centering
		\includegraphics[width=0.40\linewidth]{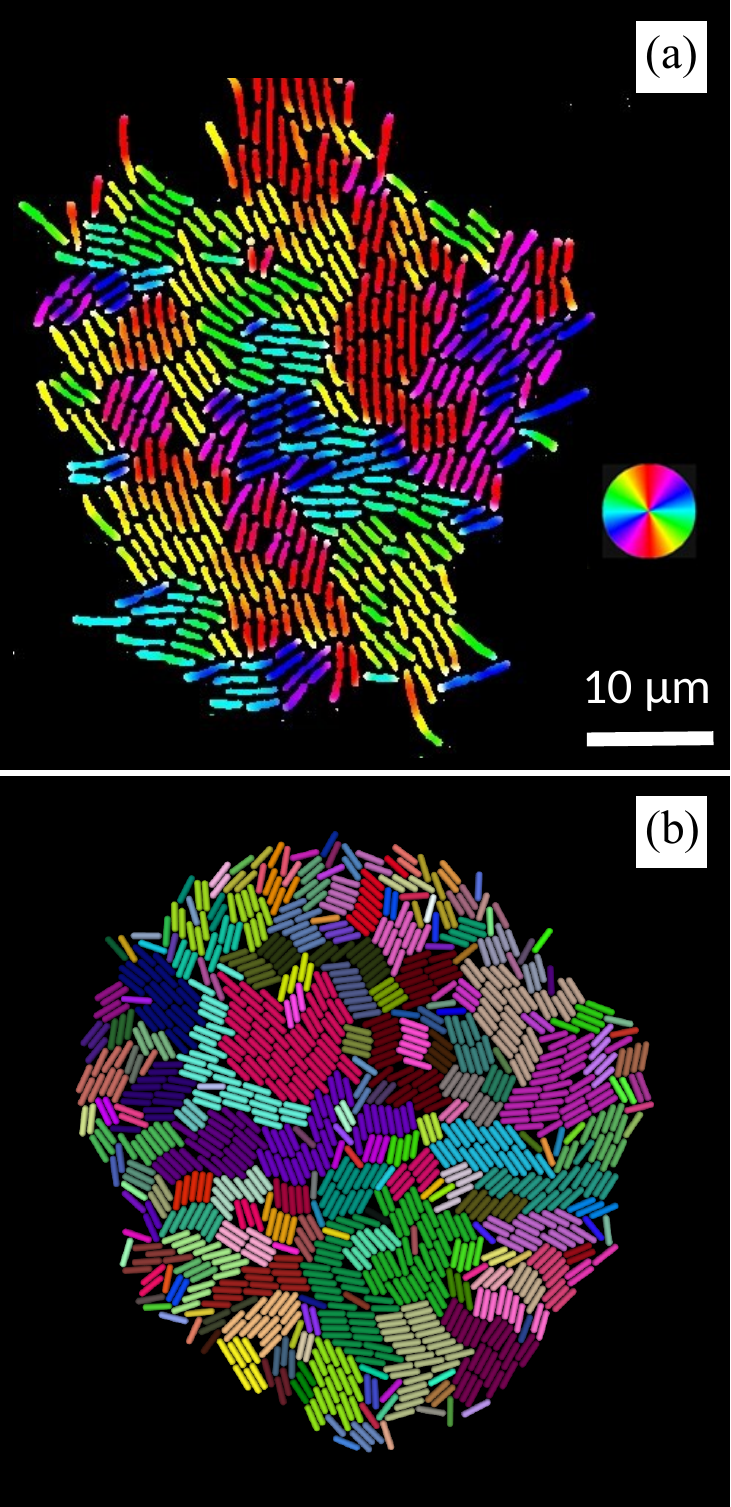}
		\caption{(Colour online) Comparison of microdomains in experimental and simulated bacterial colony. (a) Micrograph of \textit{E.~coli} (strains C600-WT) bacteria grown  on nutrient-rich agar plate.  The cells are color-coded to visualize local cell orientations. (b) Simulated colony of our rod-like cell model for a single monolayer. The color encodes membership to microdomains, identified via a density-based clustering algorithm (see text for details). Microdomains are characterized by both a large local nematic as well as local smectic order.}
		\label{fig:Comparison_exp_sim_microdomains}
	\end{figure}

	As the colony grows, the mechanical stresses mutually exerted on the cells increase, until a threshold value is met and the verticalization transition is crossed \cite{beroz2018verticalization}, resulting in the extrusion of some cells from the initial monolayer. Figure \ref{fig:4-pane_growth}(d) shows that some cells have formed the seeds of a flat, second layer, growing directly on top of the initial layer with all cells aligned parallel to the underlying substrate \cite{dhar2022self}. Figure \ref{fig:two_layer_view} shows the configuration of a colony which has developed a large second layer, visible from the central darker region in figure  \ref{fig:two_layer_view}(a) and from the edge-view in figure \ref{fig:two_layer_view}(b).

	
	Close inspection of the colonies reveals that neighboring cells exhibit a large degree of local alignment. Microdomains, best captured in bacterial monolayers, are defined as clusters of cells that display a large local nematic order parameter \cite{you2018geometry}. Figure \ref{fig:Comparison_exp_sim_microdomains}(a) shows a micrograph of bacterial colony captured experimentally, using \textit{E.~coli} (strain C600-WT) bacteria grown on nutrient-rich agar surface. The non-motile strain of \textit{E.~coli} bacteria were grown at around 30\textcelsius, showing a doubling time of approximately 40 minutes. The duration of our experiments were short enough (up to a few hours), ensuring that the expanding monolayers remained nutrient-replete throughout the observation time-scale. The growth of a single bacterium into bacterial monolayers was imaged using time-lapse microscopy (phase contrast mode), acquiring images at 5 minute intervals. Initially, the colony expanded as a monolayer (2D), spanning multiple generations, before extruding into the third dimension to reach a complex 3D structure. The color coding shows that a large degree of local nematic order is present in the monolayer system, and the local orientation can be used to define microdomains. As can be observed in Figure \ref{fig:Comparison_exp_sim_microdomains}(a, b), the intersection of three or more nematic microdomains (i.e., intersection of three or more colored clusters), typically represents topological defect sites \cite{ilnytskyi2014topological}, the dynamics of which have been discussed in detail elsewhere \cite{sengupta2020topology, dhar2022self}.

	\begin{figure}[!t]
		\centering
		\includegraphics[width=0.6\linewidth]{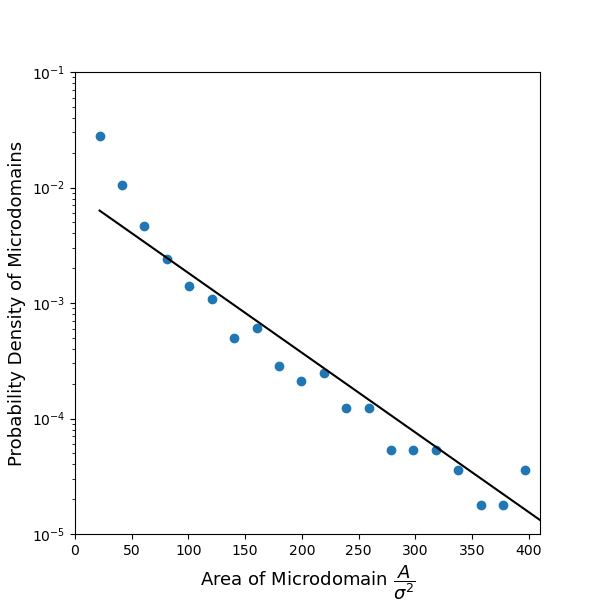}
		\caption{(Colour online) Probability distribution of the microdomains' area in mature 2D colonies before the first verticalization event. Data are averages over 39 independent simulations. The solid line is an exponential fit to the data.}
		\label{fig:Probability_Density_Histogram}
	\end{figure}

	In order to investigate more closely the presence of microdomains, we employ a density-based clustering algorithm. We consider two cells to belong to the same cluster if the angle between their orientations $\theta_{ij}<5^\circ$     and their center-of-mass distance $r_{ij}<  {0.75  l_s }$. Using these metrics, we use the density-based spatial clustering of applications with noise (DBSCAN) algorithm \cite{ester1996density}. 
	Figure \ref{fig:Comparison_exp_sim_microdomains}(b) shows the results of the DBSCAN cluster analysis on a representative model colony. Similarly to the experimental results, there is a broad distribution of microdomains, differing in size and detailed morphology. Both simulated and experimental microdomains 
	display a large local nematic order and also large local smectic order \cite{paget2023complex,paget2022smectic} (though the latter is less pronounced in the experiments,  due to the intrinsic biological variability of the cell properties). 
	
	Figure \ref{fig:Probability_Density_Histogram} shows the probability distribution of the microdomains' area $P(A)$ resulting from the clustering analysis described above and averaged over 39 independent simulations.  Our simulations show an exponential distribution $P(A)\sim \re^{-A/A^*}$ in agreement with experimental measurements \cite{you2018geometry}; furthermore, our estimate for the characteristic domain area $A^*\approx 63 \sigma^2$, which implies that the typical size of these microdomains is $l^*\approx 8 \sigma$, which also agrees with the measurement \cite{you2018geometry}.

	In summary, we have discussed a model for proliferating bacterial colonies where rod-like cells grow and divide on a substrate. The model includes steric repulsions among cells and substrate, and also the short-range attractive forces induced by the EPS matrix. The simulated colonies initially grow as a 2D layer, and after a verticalization event, form a second layer where cells are also aligned parallel to the substrate. The comparison of our simulation results with experiments on \textit{E. coli} cells shows a good match of the morphological features of the colonies. A cluster analysis of the microdomains reveals an exponential distribution of microdomain sizes, in agreement with recent experiments \cite{you2018geometry}. 
	It would be interesting to explore the statistics of the mechanical forces underpinning the formation, growth, and splitting of microdomains across the multiple layers of the bacterial colony. Specifically, the interplay of geometry and topological defects at the boundaries of microdomains with the dynamical forces shaping those domains is the primary candidate for the mechanism underpinning the development of the colony.

	\section*{Acknowledgements}
	
	This work was supported by the UK Engineering and Physical Sciences Research Council (EPSRC) grant EP/S515140/1 for Loughborough University National Productivity Investment Fund (NPIF) 2018.
	J.C. and M.G.M.
	gratefully acknowledge EPSRC grant EP/W522569/1.
	 A.S. thanks the Institute for Advanced Studies, University of Luxembourg (AUDACITY Grant: IAS-20/CAMEOS) and the Luxembourg National Research Fund's ATTRACT Investigator Grant (Grant no. A17/MS/11572821/MBRACE) and CORE Grant (C19/MS/13719464/TOPOFLUME/Sengupta) for supporting this work.

	\bibliographystyle{cmpj}
	\bibliography{biofilm}
	
\ukrainianpart

\title{До реалістичного моделювання багатошарових бактеріальних колоній}
\author[М. Т. Хан, Я. Камманн, А. Сенгупта, Е. Ренці, М. Г. Мацца]{М. Т. Хан\refaddr{label1},
	Я. Камманн\refaddr{label1},
	А. Сенгупта\refaddr{label2,label3},
	Е. Ренці\refaddr{label4},
	М. Г. Мацца\refaddr{label1}
}
\addresses{
	\addr{label1} Міждисциплінарний центр математичного моделювання та кафедра математичних наук, Університет Лафборо, Лафборо, Лестершир LE11 3TU, Великобританія
	\addr{label2} Група фізики живої речовини, Факультет фізики та матеріалознавства, Люксембурзький університет, 162 A, Авеню де ля Фаянсері, L-1511 Люксембург, Люксембург
	\addr{label3} Інститут перспективних досліджень, Люксембурзький університет, 2 Авеню де л'Універсіте, L-4365 Еш-сюр-Альзетт, Люксембург
	\addr{label4} Факультет математики, фізики та електротехніки, Нортумбрійський університет, Ньюкасл-апон-Тайн, NE1 8ST, Великобританія
}

\makeukrtitle

\begin{abstract}
	Бактерії продуктивно колонізують різноманітні поверхні в широкому діапазоні умов навколишнього середовища. Вони демонструють захоплюючі приклади самоорганізації на різних масштабах. Незважаючи на те, що роль механічних сил у колективній поведінці бактеріальних колоній вивчається протягом тривалого часу, вона ще не повністю зрозуміла. У даній роботі ми розглядаємо модель росту паличкоподібних бактерій, таких як \textit{Escherichia coli}, засновану виключно на механічних силах. Ми виконуємо моделювання колонії методом сильно демпфованої молекулярної динаміки, починаючи з кількох клітин, які контактують із поверхнею. У міру зростання колонії мікроскупчення сильно впорядкованих клітин ростуть і розмножуються. Наша модель відслідковує як початковий ріст бактеріальної колонії, так і демонструє багато\-обiцяючi ознаки фіксації експериментально спостережуваного переходу до багатошарових колоній протягом більш тривалого періоду часу. Ми порівнюємо наші результати з експериментами на клітинах \textit{E.~coli} та аналізуємо статистику мікродоменів.
	\keywords бактерії, активне середовище, орієнтаційний порядок, геометрія, механіка, перехід з одно- до багатошарового стану
\end{abstract}

\end{document}